\documentclass[aps,prd,10pt,notitlepage,nofootinbib]{revtex4-1}

\usepackage[utf8]{inputenc}
\usepackage{amsmath,amssymb,amsfonts}
\usepackage{newtxtext,newtxmath}
\usepackage{bbold}
\usepackage{bm}
\usepackage{graphicx}
\usepackage[usenames,dvipsnames]{xcolor}
\usepackage{color}
\usepackage[colorlinks=true,linkcolor=cyan,urlcolor=cyan,citecolor=cyan]{hyperref}
\usepackage{slashed}
\usepackage[english]{babel}
\usepackage{dcolumn}
\usepackage{pifont}
\usepackage{dsfont,mathrsfs}
\usepackage{cancel}
\usepackage{bigints}
\usepackage{accents}
\usepackage{soul}
\usepackage{multirow}

\usepackage{amsmath, amssymb}
\usepackage{bbold}
\usepackage{makecell}
\usepackage{setspace}
\usepackage{titlesec}
\usepackage{ulem} 
\usepackage{CJKutf8}
\usepackage{orcidlink}

\allowdisplaybreaks

\newcommand{\nn}{\nonumber}
\newcommand{\ket}[1]{\left|#1\right\rangle}
\newcommand{\bra}[1]{\left\langle#1\right|}

\newcommand{\FB}[1]{\left(#1\right)}

\newcommand{\SB}[1]{\left\{#1\right\}}
\newcommand{\TB}[1]{\left[#1\right]}

\newcommand{\munu}{{\mu\nu}}

\newcommand{\util}{\widetilde{u}}
\newcommand{\btil}{\widetilde{b}}

\newcommand{\Pcal}{\mathcal{P}}

\begin{document}
	\title{Higher-order flow coefficients of source-level dilepton emission in a magnetized hadronic medium \footnote{Defu Hou's research is supported in part by the National Key Research and Development Program of China under Contract No. 2022YFA1604900. Additionally, Defu Hou receives partial support from the National Natural Science Foundation of China (NSFC) under Grant No.12435009 and No. 12275104. Rajkumar Mondal is supported by Central China Normal University and by research grant from Prof. Defu Hou.}}
	
	\author{Rajkumar Mondal\orcidlink{0000-0002-1446-6560}$^{a}$}
	\email{rajkumarmondal.phy@gmail.com / rajkumarmondal.phy@zohomail.in  \textcolor{black}{(corresponding author)}}
	
	\author{Defu Hou \begin{CJK}{UTF8}{gbsn}(侯德富)\end{CJK} $^{a}$}
	\email{houdf@ccnu.edu.cn}

	\affiliation{$^a$Institute of Particle Physics and Key Laboratory of Quark and Lepton Physics (MOS),\\Central China Normal University, Wuhan 430079, China}


\begin{abstract}
The study of dilepton emission from hot hadronic matter provides a unique probe of the properties of strongly interacting medium created in heavy-ion collisions. In non-central collisions, the presence of magnetic fields can induce anisotropic features in the emission spectrum.  While the impact of a magnetic field on the dilepton emission rate has extensively studied, the detailed higher-order azimuthal anisotropies of the emission rate--characterized by flow coefficients beyond elliptic flow--
remain an open question, particularly in the low invariant-mass region where magnetic-field-induced medium effects are expected to be most pronounced. Here, we investigate higher-order azimuthal anisotropies of the source-level thermal dilepton emission from a magnetized hot hadronic medium. Our results reveal a continuous dilepton spectrum with strong Landau-cut contributions at low invariant masses due to the background magnetic field. The emission rate exhibits significant azimuthal-angle dependence in this region, characterized by source-level flow coefficients $v_{2,4,6}$. The odd-order coefficients vanish due to symmetry. The elliptic flow coefficient $(v_2)$ is positive and exhibits an oscillatory structure at low invariant masses--driven by Landau-level quantization of pions, but negligible at higher masses. Similar behavior is observed for the higher-order coefficients $v_4$ and $v_6$ with notable structures at low masses and negligible values at higher masses. The results highlight dileptons as sensitive probes of magnetic-field effects in heavy-ion collisions, offering new avenues to constrain the strength of magnetic fields and unravel the properties and dynamics of hot hadronic matter.

\end{abstract}

\maketitle

\section{Introduction}
Strongly interacting hot and dense nuclear matter, composed of quarks or hadrons, is anticipated to be created at relativistic heavy-ion collisions (HICs) experiments at facilities like the Relativistic Heavy Ion Collider (RHIC) at Brookhaven National Laboratory and the Large Hadron Collider (LHC) at CERN. Such matter may also exist in the cores of compact astrophysical objects, such as neutron stars. The collision of two nuclei at ultra-relativistic energies leads to the liberation of the fundamental constituents of the nucleons forming a deconfined state of quarks and gluons, which is known as the quark-gluon plasma (QGP) in local thermal equilibrium~\cite{Wong:1995jf,Sarkar:2010zza,Florkowski:2010zz,Satz:2012zza}. The matter created in HICs undergoes rapid cooling due to expansion driven by its own pressure gradient, undergoing various stages of evolution. However, direct observation of this process is strongly hindered due to its transient ($\sim$ few fm/c) nature of the process. Consequently, researchers must rely on indirect probes and observables such as spectra of electromagnetic probes (photons and dileptons), heavy quark production, quarkonia suppression, jet energy loss, collective flow, $J/\Psi$ suppression etc to extract microscopic as well as bulk properties of the create matter. Among these, electromagnetic probes (photons and dileptons) are the most effective probes for characterizing the space-time history of the collision~\cite{McLerran:1984ay,Kajantie:1986dh,Gale:1988vv,Weldon:1990iw,Ruuskanen:1990hx,Ruuskanen:1991au,Alam:1996fd,Alam:1999sc}. Owing to their large mean free paths, they typically leave the system without much interactions and therefore are expected to carry undistorted information about the stage of their production. It is important to note that experimentally measured anisotropic flow coefficients are extracted from the final-state particle multiplicity distributions and arise from the space-time evolution of the medium, including hydrodynamic expansion and event-by-event fluctuations~\cite{Dion:2011pp}. In contrast, theoretical calculations of electromagnetic emission typically provide the local production rate, which must be integrated over the full evolution of the system to obtain observable quantities.
In the present work, source-level	refers to the local thermal dilepton emission rate from the hadronic source considered here, before incorporating the space-time evolution of the medium and final-state effects. The corresponding $v_n$ therefore characterize the intrinsic azimuthal anisotropy of this	local emission source and should not be identified with experimentally	measured final-state flow coefficients.
In the QGP, quarks and antiquarks can annihilate to form a virtual photon, which subsequently decays into a lepton-antilepton pair. The dilepton emission rate from this process has been widely studied in Refs.~\cite{Kajantie:1986dh,Ruuskanen:1990hx,Ruuskanen:1991au}. In heavy-ion collisions, however, several other thermal and non-thermal sources also produce dileptons and create a sizable background~\cite{Ruuskanen:1991au,Wong:1995jf,Alam:1996fd,Cassing:1999es}. Among these, the Drell--Yan process is one of the most important and is well understood within perturbative QCD~\cite{Craigie:1978bp,Grosso-Pilcher:1986iaq,CTEQ:1993hwr,Wong:1995jf,Alam:1996fd}. 
Dileptons are also produced from the decays of hadronic resonances such as $\pi^0$, $\rho$, $\omega$ and $J/\psi$ and their yields can be estimated experimentally through invariant-mass analysis~\cite{Wong:1995jf}. However, separating the photons and dileptons originating from the hadronic medium, formed after the phase transition or crossover, is challenging. Therefore, a reliable theoretical estimate of the photon and dilepton yields from a hot and dense hadronic medium, including the possible modification of hadronic properties below the critical temperature, is crucial for identifying electromagnetic signals from the QGP. Extensive studies have shown that the dilepton emission rate from the hadronic phase is significantly modified in the low invariant mass region~\cite{McLerran:1984ay,Weldon:1990iw,Gale:1990pn,Rapp:1999ej,Mallik:2016anp}.

It is conjectured that a very strong magnetic field is produced in non-central or asymmetric HICs  at the RHIC and LHC. The estimated value of the magnetic field in the HICs is around $10^{15-18}$ Gauss~\cite{Tuchin:2013ie,Kharzeev:2007jp,Skokov:2009qp,Tuchin:2013apa} generated by the rapid movement of electrically charged spectators in the early stages of the collision.  Although it experiences rapid decay within a few fm/c, the finite electrical conductivity ($\sim$ a few MeV) of the medium can possibly delay the decay process sufficiently allowing a nonzero magnetic field to persist even during the subsequent hadronic phase following the phase transition or crossover from the QGP~\cite{Gursoy:2014aka,Inghirami:2016iru,Kalikotay:2020snc}. 
The strength of these magnetic fields is comparable to the typical energy scale of quantum chromodynamics (QCD), affecting various microscopic and macroscopic properties of strongly interacting matter (Refs.~\cite{Friman:2011zz,Miransky:2015ava,Kharzeev:2013jha} for reviews), leading to exotic phenomena such as the Chiral Magnetic Effect (CME), Chiral Vortical Effect (CVE), Magnetic Catalysis (MC), and Inverse Magnetic Catalysis (IMC). The effects of a magnetic field on the transport properties of hadronic matter, including shear and bulk viscosities as well as electrical conductivity, have been extensively studied in Refs.~\cite{Kadam:2014xka,Das:2019pqd,Dash:2020vxk,Ghosh:2022xtv,Das:2019wjg,Kalikotay:2020snc,Das:2020beh}. In addition, the in-medium properties of vector mesons, such as the $\phi$ meson, have also been investigated under a strong magnetic field to understand modifications of hadronic matter in extreme conditions~\cite{Chahal:2024oib}.

The presence of a magnetic field significantly modifies the emission of electromagnetic probes. These fields can induce anisotropies in the emission rates of electromagnetic probes. Understanding such effects at the level of the local production rate is important for disentangling different physical mechanisms contributing to the overall observed anisotropy. Dilepton production from the QGP medium under such conditions has been extensively explored in the literature by many authors~\cite{Tuchin:2012mf,Tuchin:2013bda,Sadooghi:2016jyf,Bandyopadhyay:2016fyd,Bandyopadhyay:2017raf,Ghosh:2018xhh,Islam:2018sog,Ghosh:2020xwp,Hattori:2020htm,Chaudhuri:2021skc,Wang:2022jxx,Das:2021fma, Dwibedi:2025xho, Panda:2025yxw, Castano-Yepes:2025zae, Ayala:2025jpr, Mustafa:2025uad} employing various approximations. Several studies~\cite{Wang:2020dsr,Wang:2022jxx,Ayala:2024ucr,Wang:2023fst} have investigated anisotropies in electromagnetic emission induced by magnetic fields. However, these analyses often focus on specific observables or particular phases of the medium and a systematic understanding of the angular structure of dilepton emission remains incomplete. For example, Ref.~\cite{Wang:2020dsr} investigates the ellipticity of photon emission from a hot magnetized QCD plasma, while Ref.~\cite{Wang:2022jxx} determines the ellipticity parameter  $(v_2)$ for dileptons emission from a magnetized quark-gluon-plasma showing that $v_2$ tends to be positive and large at large transverse momenta. Ref.~\cite{Ayala:2024ucr} shows that the most favored direction of photon emission is the plane transverse to the magnetic field, which is consistent with a positive contribution to the elliptic flow coefficient. Higher-order anisotropy coefficients in photon and dilepton emission from QGP are discussed in Ref.~\cite{Wang:2023fst}. As the system expands and cools from the QGP state, hadronic matter forms and serves as a significant source of dilepton production, particularly in the low invariant mass region. 
Recent studies in Refs.~\cite{Mondal:2023vzx,Mondal:2023ypq} analyze dilepton production and its elliptic flow coefficient  $(v_2)$ from a magnetized hadronic medium in terms of the spectral function of $\rho^0$, evaluated from the electromagnetic vector current correlation function using the real-time formalism of thermal field theory. 

In the present work, we investigate higher-order angular anisotropies of the source-level dilepton emission in a magnetized hadronic medium. The coefficients defined in this work quantify the anisotropy of the local production rate and provide a baseline for future studies that incorporate the full space-time evolution of the medium. To the best of our knowledge, such a study has not yet been performed.

The article is organized as follows: Section \ref{Formalism} presents the general formalism for anisotropy in the dilepton production rate, expressed in terms of the spectral function in a thermomagnetic background. Section \ref{Numerical} discusses the numerical results, and Section \ref{SC} provides a summary and conclusions.


\section{Formalism}\label{Formalism}

\subsection{Dilepton emission rate from hadronic medium}\label{DPR_From_Hadron}
In a hadronic medium, dileptons are generated through hadronic resonances, where a virtual photon emitted during a hadronic transition subsequently decays into a lepton–antilepton pair. The method for evaluating dilepton production from a hot hadronic medium is well established in earlier studies~\cite{Kajantie:1986dh,Gale:1988vv,Gale:1987ki,Alam:1999sc,Ruuskanen:1989tp,Mallik:2016anp,Laine:2016hma}. But, for the sake of completeness, we only summarize the main steps required for our analysis following the approach of Ref.~\cite{Mallik:2016anp}.

We consider a transition from an initial thermal state
$\ket{i}=\ket{I(\mathcal{P}_I)}$ with momentum $\mathcal{P}_I$ to a final state $\ket{f}=\ket{H(\mathcal{P}_H);\,l(p),\bar l(p')}$ consisting of hadrons with momentum $\mathcal{P}_H$ and emitted dilepton carries four-momentum $q^\mu=p^\mu+p'^\mu$. Such transition probability is determined from the squared matrix element $|\bra{f}\hat S\ket{i}|^2$, where the scattering matrix operator is expressed as
\begin{equation}
	\hat S=\tau_c\exp\!\left[i\int d^4x\,\mathcal{L}_{\rm int}(x)\right],
\end{equation}
in which $\tau_c$ denotes the time-ordering operator and the interaction Lagrangian density relevant for the electromagnetic processes is~\cite{Mallik:2016anp}
\begin{equation}
	\mathcal{L}_{\rm int}=-e\left[J_\mu^l(x)+J_\mu^h(x)\right]\mathcal{A}^\mu(x)
\end{equation}
where $J_\mu^l(x)=e\bar{\psi}(x)\gamma_\mu\psi(x)$ and $J_\mu^h(x)$ represent the conserved leptonic and hadronic electromagnetic currents, respectively, coupled through the photon field $\mathcal{A}^\mu(x)$. 

\begin{figure}[h]
	\includegraphics[scale=.4]{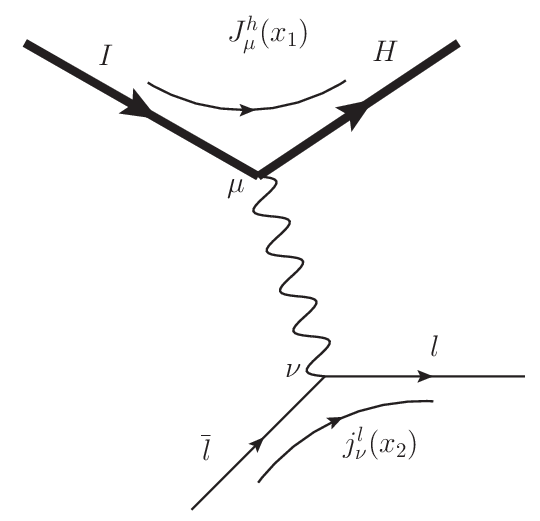}	
	\caption{Feynman diagram for dilepton production from hadronic medium.}\label{Fey.Diag}
\end{figure}

Note that electromagnetic probes interact weakly with the medium ($e \ll 1$), while the medium itself remains strongly interacting. Since the electromagnetic coupling is perturbatively small compared to the strong interaction scale, electromagnetic emission can be treated as a perturbation on top of an equilibrated strongly interacting medium. This hierarchy of scales justifies expanding the scattering operator to leading non-vanishing order in the electromagnetic interaction. In addition, owing to their large mean free paths, electromagnetic probes escape the medium with minimal final-state interaction.
Now, expanding the $S$-matrix to second order in the interaction,
	\begin{equation}
		\hat S^{(2)} = \frac{(-i)^2}{2!}\int d^4x_1 d^4x_2\, \tau_c\{\mathcal{L}_{\rm int}(x_1)\mathcal{L}_{\rm int}(x_2)\},
	\end{equation}
the first non-trivial contribution arises from the mixed hadronic--leptonic term. The corresponding Feynman diagram is shown in Fig.~\ref{Fey.Diag}; disconnected diagrams do not contribute to this process and are therefore omitted. 	Thus, the transition amplitude becomes~\cite{Mallik:2016anp}
\begin{equation}
	\langle f|\hat S|i\rangle
		=-e^2\!\int d^4x_1 d^4x_2
		\langle H|J_\mu^h(x_1)|I\rangle
		\langle l,\bar l|J_\nu^l(x_2)|0\rangle
		\Delta_F^{\mu\nu}(x_1-x_2).
		\label{fsi}
\end{equation}
The strong interaction dynamics are entirely contained in the hadronic matrix element $\langle H|J^h_\mu(x_1)|I\rangle$, while the leptonic matrix element $\langle l,\bar{l}|J^l_\nu(x_2)|0\rangle$ can be evaluated explicitly in terms of free Dirac spinors. The contraction of the photon fields gives rise to the free photon propagator $\Delta_F^{\mu\nu}(x_1-x_2)$.
Performing the space-time integration over the leptonic vertex leads to a four-momentum conserving delta function, which fixes the internal photon momentum to be equal to the total dilepton momentum $q = p + p'$. As a result, the amplitude reduces to a single space-time integral over the hadronic current, multiplied by the leptonic current factor.
Finally, using translational invariance, the current operator can be shifted to the origin, which extracts the overall energy–momentum conserving delta function $(2\pi)^4 \delta^{(4)}(P_I - P_F - q)$, leaving the matrix element of the current evaluated at zero space-time coordinate.	
After straightforward calculations, the squared matrix element per unit four-volume becomes~\cite{Mallik:2016anp}
\begin{equation}
	|\bra{f}\hat S\ket{i}|^2
	={(2\pi)^4}\delta^{(4)}(\mathcal{P}_I-\mathcal{P}_H-q)\,
	\frac{e^4}{q^4}\,
	l^{\mu\nu}\,
	\bra{H}J_\mu^h(0)\ket{I}\bra{I}J_\nu^h(0)\ket{H},
	\label{SqAmp}
\end{equation}
where the leptonic tensor $l^{\mu\nu}$ arises from the spin sums over the outgoing lepton pair, given by
\begin{equation}
	l^{\mu\nu}	=	4\!\left[p^\mu p'^\nu+p^\nu p'^\mu-(p\!\cdot\!p'+m_{\rm lep}^2)g^{\mu\nu}\right]
\end{equation}
with $m_{\rm lep}$ as leptonic mass.
Assuming local thermal equilibrium, we sum over final hadronic states and average over initial states with the Boltzmann weight $Z^{-1}e^{-\beta\mathcal{P}_I^0}$ \textcolor{black}{(where $\beta=1/T$ is inverse temperature and $Z$ is partition function)}. Using the energy--momentum conserving delta function, $\delta^{(4)}(\mathcal{P}_I-\mathcal{P}_H-q)$, the initial energy is written as $\mathcal{P}_I^0=\mathcal{P}_H^0+q_0$, yielding a factor $e^{-\beta q_0}$. The delta function is then expressed in its Fourier representation,	$\delta^{(4)}(p)=\int\frac{d^4x}{(2\pi)^4}\,e^{ip\cdot x}$	which cancels the prefactor $(2\pi)^4$ in Eq.~(\ref{SqAmp}). Using translational invariance,
	$J_\mu^h(x)=e^{i\hat P\cdot x}J_\mu^h(0)e^{-i\hat P\cdot x}$ and inserting a complete set of hadronic states, the product of matrix elements can be rewritten as $\sum_{I,H}Z^{-1}e^{-\beta\mathcal{P}_I^0}
	\bra{H}J_\mu^h(0)\ket{I}\bra{I}J_\nu^h(0)\ket{H}
	=\int d^4x\,e^{iq\cdot x}\,\langle J_\mu^h(x)J_\nu^h(0)\rangle,$
	which leads to~\cite{Mallik:2016anp}
\begin{equation}
	\sum_{H,I} Z^{-1} e^{-\beta\mathcal{P}_I^0}
	|\bra{f}\hat S\ket{i}|^2
	=
	\frac{e^4}{q^4} e^{-\beta q_0}
	l^{\mu\nu}
	\int d^4x\,e^{iq\cdot x}
	\langle J_\mu^h(x)J_\nu^h(0)\rangle.
\end{equation}
The dilepton multiplicity from thermal hadronic medium is then obtained by summing over the final states and averaging over the initial states, given by the relation:
\begin{equation}
	N=\sum_{\rm spin}\sum_{I,H}
	Z^{-1}e^{-\beta\mathcal{P}_I^0}
	|\bra{f}\hat S\ket{i}|^2.
\end{equation}
Introducing $\int d^4q\,\delta^{(4)}(q-p-p')=1$ in the above equation and performing the phase-space integrations, the total differential multiplicity is~\cite{Mallik:2016anp} 
\begin{equation}
	\frac{dN}{d^4x~d^4q}=\frac{e^4}{q^4}~e^{-\beta q_0} ~L_{\mu\nu}~M^{+\mu\nu}\label{Eq.DPR1}
\end{equation}
where $M^{+\mu\nu}$ denotes Fourier transform of the two-point vector current correlator and $L^{\mu\nu}$ represents for leptonic contribution, given by
\begin{align}	
&M^{+}_{\mu\nu}=\int d^4x \, e^{iq\cdot x}\langle J_\mu^h(x)J_\nu^h(0)\rangle,\\
&L^{\mu\nu}=\frac{1}{(2\pi)^6}\frac{2\pi}{3}
	\!\left(1+\frac{2m_{\rm lep}^2}{q^2}\right)
	\!\sqrt{1-\frac{4m_{\rm lep}^2}{q^2}}
	(q^\mu q^\nu-q^2 g^{\mu\nu}).
\end{align}
From Eq.~\eqref{Eq.DPR1}, the differential dilepton emission rate with four-momenta $q^\mu$ can be rewritten as~\cite{Mallik:2016anp}
\begin{equation}
	\frac{dN}{d^4x\,d^4q}
	=\frac{\alpha^2}{6\pi^3 q^2}e^{-\beta q_0}	L(q^2) \big[-g_{\mu\nu}M^{+\mu\nu}(q)\big],
	\label{DPR1}
\end{equation}
where $\alpha$ as the fine-structure constant and $L(q^2)=\sqrt{1-\frac{4m_{\rm lep}^2}{q^2}}\FB{1+\frac{2m_{\rm lep}^2}{q^2}}$. We note that the above expression is valid for $q^2>4m_{\rm lep}^2$, below which dilepton production is kinematically forbidden.

In the real-time formalism of thermal field theory, the correlator assumes a $2\times2$
matrix structure on account of the shape of the contour in complex time plane. Fourier transform of the time-ordered two-point  is
\begin{equation}
	M_{ab}^{\mu\nu}
	=	i\int d^4x\,e^{iq\cdot x}
	\langle  \tau_c[J_h^\mu(x)J_h^\nu(0)]\rangle_{ab}\label{Eq.MmuNuan}
\end{equation}
where $\tau_c$ is time-ordering operator for the contour $c$ and the thermal indices $a, b \in \{1,2\}$ indicate that the two points can chosen on either of the two horizontal segments of the time contour $c$.  It can be diagonalized as
\begin{equation}
	M^{\mu\nu}=
	U
	\begin{pmatrix}
		\overline{M}^{\mu\nu} & 0 \\
		0 & -\overline{M}^{*\mu\nu}
	\end{pmatrix}
	U,\label{Mmunudia}
\end{equation}
with $U =
\begin{pmatrix}
	\sqrt{n+1} & \sqrt{n} \\
	\sqrt{n}   & \sqrt{n+1}
\end{pmatrix}$ containing the thermal distribution function
$n=(e^{\beta|q_0|}-1)^{-1}$. The diagonal element $\overline{M}^{\mu\nu}$ in Eq.~\eqref{Mmunudia} is an analytic function and it is related to the $11$ component via
\begin{eqnarray}
	\text{Re}\,\overline{M}^{\mu\nu}(q)
	&=&
	\text{Re}\,M_{11}^{\mu\nu}(q),\\
	\text{Im}\,\overline{M}^{\mu\nu}(q)
	&=&\tanh\!\left(\frac{|q_0|}{2T}\right)
	\text{Im}\,M_{11}^{\mu\nu}(q).
\end{eqnarray}
Using the spectral representation, one finds 
\begin{equation}
	M^{+\mu\nu}(q)
	=
	\frac{2e^{\beta q_0}}{e^{\beta q_0}-1}
	\text{Im}\,\overline{M}^{\mu\nu}(q)
	=\epsilon(q^0)
	\frac{2e^{\beta q_0}}{e^{\beta q_0}+1}
	\text{Im}\,{M}^{\mu\nu}(q)\label{Mmunuplus}
\end{equation}
where $\epsilon(q^0)$ is the sign function. Substitution Eq.~\eqref{Mmunuplus} into Eq.~\eqref{DPR1}, we obtain the dilepton production rate (DPR) as 
\begin{equation}
	\frac{dN}{d^4x\,d^4q}
	=
	\frac{\alpha^2}{3\pi^3 q^2}
	\frac{1}{e^{\beta q_0}-1}
	L(q^2)
	\big[-g_{\mu\nu}\text{Im}\,\overline{M}^{\mu\nu}(q)\big].
	\label{DPR1.1}
\end{equation}
To compute the DPR, we require an explicit form of the local hadronic current. Within the Vector Meson Dominance (VMD)~\cite{Sakurai:1960ju,Bhaduri:1988gc,Mallik:2016anp} framework, the conserved hadronic current can be expressed as
\begin{equation}
	J_h^\mu(x) = J_\rho^\mu(x) = F_\rho\, m_\rho\, \rho^\mu(x),\label{Eq.JHadronicCurrent}
\end{equation}
where $\rho^\mu(x)$ denotes the Heisenberg field corresponding to the $\rho^0$--meson, the coupling $F_\rho = 156$ MeV is fixed from the decay rate $\Gamma_{\rho^0 \to e^+ e^-} = 7.0~\text{keV}$~\cite{Mallik:2016anp}. Substituting Eq.~\eqref{Eq.JHadronicCurrent} into Eq.~\eqref{Eq.MmuNuan} and applying Wick's theorem, we can arrive at
\begin{eqnarray}
	\text{Im}~M_{11}^{\munu}(q)= F^2_\rho m^2_\rho \text{Im}~D^{\munu}_{11}(q)	\label{ImM11munu}
\end{eqnarray} 
where $D^{\munu}_{11}(q)$ is 11-component of exact thermal $\rho_0$ propagator. Finally, using Eq.~\eqref{Mmunuplus}, Eq.~\eqref{DPR1.1} and Eq.~\eqref{ImM11munu}, the dilepton production rate reads~\cite{Mallik:2016anp}
\begin{equation}
	\frac{dN}{d^4x\,d^4q}
	=
	\frac{\alpha^2}{\pi^3 q^2}
	f_{\rm BE}(q_0)
	L(q^2)
	F_\rho^2 m_\rho^2
	\mathcal{A}(q;T),
	\label{DPR}
\end{equation}
where $f_{\rm BE}(x)=\left(e^{x/T}-1\right)^{-1}$ denotes the Bose--Einstein distribution function and 
$\mathcal{A}(q;T)=-\frac{1}{3}g^{\mu\nu}\,\mathrm{Im}\,\overline{D}_{\mu\nu}(q)$ represents the in-medium $\rho^0$ spectral function with $\overline{D}_{\mu\nu}(q)$ being the full $\rho^0$-meson propagator in absence of a magnetic field.

In a thermomagnetic background, magnetic-field effects are incorporated only in the hadronic sector through modifications of the vector-meson propagator. The leptonic tensor is treated in its vacuum form, assuming that the produced dileptons, owing to their large mean free path and weak electromagnetic interaction, escape the medium with negligible final-state interactions and are effectively detected outside the strongly magnetized region. Consequently, magnetic-field effects on the leptonic sector are neglected and all magnetic-field dependence is encoded in the in-medium $\rho^0$ propagator, leading to the modified spectral function
\begin{eqnarray}
		\mathcal{A}(q;T,eB) = -\frac{1}{3}g^{\mu\nu}\,\mathrm{Im}\,D_{\mu\nu}(q),
\end{eqnarray}
with ${D}_{\mu\nu}(q)$ being the full $\rho^0$-meson propagator in presence of a magnetic field, will be discussed in Sec.~\ref{DPRB}.

\subsection{{Flow} coefficients  for source-level dilepton emission in a background magnetic field}
\begin{figure}[h]
	\includegraphics[scale=0.7]{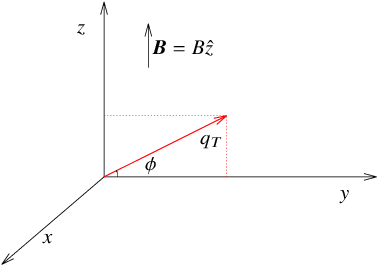}
	\caption{A schematic diagram of the coordinate frame used in this work}
	\label{Frame}
\end{figure}

The flow coefficients characterize the angular distribution of dilepton emission and magnetic-field-induced properties of the medium. In the present work, however, we emphasize anisotropy coefficients of source-level dilepton emission.
We follow the geometry shown in Fig.~\ref{Frame} to study the anisotropy of the dilepton emission rate in the presence of a background magnetic field. For simplicity, we take the magnetic field to be constant and oriented along the $z$-axis, while the beam direction is chosen along the $x$-axis. With this setup, the transverse momentum of the dilepton lies in the $y$--$z$ plane, which is perpendicular to the beam direction. The azimuthal angle $\phi$ is defined with respect to the reaction plane, i.e., the $x$--$y$ plane shown in Fig.~\ref{Frame}. 
Thus, $\phi\sim0$ corresponds to the direction within the reaction plane, which is transverse to the background magnetic field oriented along the $z$ axis.
In relativistic heavy-ion collisions, the experimentally observed anisotropic flow coefficients are extracted from the final particle multiplicity distributions. These observables are obtained by integrating the local (source-level) emission rate over the full space-time evolution of the medium. Consequently, the anisotropy coefficients defined at the level of the local rate represent intrinsic emission properties and may be modified by the subsequent space-time evolution of the medium.

In non-central heavy-ion collisions, the differential dilepton emission rate is typically expressed as a Fourier expansion in the azimuthal angle relative to the reaction plane, expressed by~\cite{Voloshin:1994mz},
\begin{equation}
	\frac{dN}{d^4xd^4q}=\frac{1}{2\pi}\frac{dN}{d^4xMdMq_Tdq_Tdy}\TB{1+\sum_{n=1}^{\infty}2v_n{\rm~cos}\SB{n(\phi-\phi_{re})}}
\end{equation}
where the coefficients $v_n$ quantify the anisotropy of the differential production rate in momentum space at a given space-time point and $\phi_{re}$ denotes the reaction-plane angle. The dilepton four-momentum is written as
\[q^\mu=\FB{q^0,\bm{q}}\equiv (q^0,q_x,q_y,q_z)\equiv(q^0,q_x,q_T\cos\phi,q_T\sin\phi)\] with $q_T=\sqrt{q_y^2+q_z^2}$ 
being the transverse momentum. The infinitesimal four-momentum element is \[d^4q=MdMq_Tdq_Td\phi dy\]
where $M=\sqrt{q^2}$ is the invariant mass and the rapidity describing motion of the particle is \[y=\dfrac{1}{2}\ln\FB{\dfrac{q_0+q_x}{q_0-q_x}}.\] 
In this work, the calculation is performed in the rest frame of the medium. In a realistic evolving system, this frame differs from the laboratory frame due to collective flow and both the dilepton momentum and the electromagnetic field undergo Lorentz transformations. For simplicity, we restrict to $q_x=0$, corresponding to vanishing longitudinal momentum in the chosen frame. Owing to rotational symmetry about the magnetic-field direction ($\hat{z}$), this choice is arbitrary and does not affect the anisotropy coefficients. This restriction simplifies the analysis but limits direct comparison with experimentally extracted observables. The anisotropy coefficients $v_n$ are then obtained as:
\begin{equation}
	v_n=\frac{{\int_{0}^{2\pi}\frac{dN}{d^4xd^4q}}{\rm~cos}(n\phi)d\phi}{{\int_{0}^{2\pi}\frac{dN}{d^4xd^4q}}d\phi}\label{Vn}.
\end{equation}
These coefficients therefore characterize the angular anisotropy of the local emission rate induced, in the present context, by the external magnetic field.
In the zero-field limit, $eB=0$, finite temperature modifies the magnitude of the thermal dilepton production rate but does not introduce a preferred azimuthal direction. Consequently, the production rate is azimuthally isotropic and the corresponding anisotropy coefficients $v_n$ vanish, providing a reference for the magnetic-field-induced source-level anisotropy studied in this work.
They should be interpreted as source-level anisotropies, which provide a baseline for more realistic calculations including hydrodynamic or transport evolution.
The denominator in Eq.~\eqref{Vn} represents the total differential dilepton emission rate. The formalism for dilepton emission from a thermal hadronic medium has been studied in the previous Sec.~\ref{DPR_From_Hadron}.

\subsection{Rho-meson spectral function in a background magnetic field}\label{DPRB}
The key quantity in the dilepton emission rate from hadronic matter is the in-medium spectral function of the $\rho^0$, which is evaluated using the VMD model in the presence of a background magnetic field~\cite{Mallik:2016anp,Gale:1990pn,Mallik:2016anp,Rapp:1999ej}.  
This spectral function corresponds to the imaginary part of the full $\rho^0$ propagator and contains the dynamics of the hadronic medium. It can be obtained by solving the Dyson-Schwinger equation, which relates the $\rho^0$ self-energy $\Pi^{\mu\nu}$ and the bare propagator $D^{\mu\nu}_0$~\cite{Mallik:2016anp,Bellac:2011kqa}:
\begin{equation}
	D^{\mu\nu}=D^{\mu\nu}_{0}-D^{\mu\beta}_{0}\Pi_{\beta\alpha}D^{\alpha\nu}\label{Dyson}
\end{equation} 
where the bare propagator is  
\[
D^{\mu\nu}_{0}(q) = \frac{-1}{q^2 - m_\rho^2 + i\epsilon} \left(-g^{\mu\nu} + \frac{q^\mu q^\nu}{m_\rho^2} \right).
\]  
We perform a Lorentz decomposition of the $\rho^0$ self-energy in the presence of a magnetic field to obtain the spectral function $\mathcal{A}(q;T,eB)$~\cite{Mondal:2023vzx}:
\begin{equation}
	\Pi^{\mu\nu}(T,eB) = \Pi_1 \Pcal_1^{\mu\nu} + \Pi_2 \Pcal_2^{\mu\nu} + \Pi_3 \Pcal_3^{\mu\nu} + \Pi_4 \Pcal_4^{\mu\nu}. \label{eq.PiB.dec}
\end{equation}
Here, $\Pcal_i^{\mu\nu}$ $(i=1,\ldots,4)$ are the basis tensors constructed from the four-vectors $q^\mu$, $u^\mu$, and $b^\mu$ available in the magnetized thermal medium. The medium four-velocity $u^\mu$ and the magnetic-field four-vector $b^\mu$ introduce preferred directions in addition to $q^\mu$. By constructing the components of $u^\mu$ and $b^\mu$ transverse to $q^\mu$, a convenient tensor basis $\Pcal_i^{\mu\nu}$ can be formed, leading to the Lorentz decomposition of the $\rho^0$ self-energy in	Eq.~\eqref{eq.PiB.dec}. The corresponding scalar form factors $\Pi_i$ $(i=1,\ldots,4)$ contain the dynamical information of the thermal medium and the external magnetic field. They are given by
\begin{eqnarray}
	\Pcal_1^{\mu\nu} &=& \frac{\util^\mu \util^\nu}{\util^2}, \quad
	\Pcal_2^{\mu\nu} = g^{\mu\nu} - \frac{q^\mu q^\nu}{q^2} - \frac{\util^\mu \util^\nu}{\util^2} - \frac{\btil^\mu \btil^\nu}{\btil^2}, \nonumber \\
	\Pcal_3^{\mu\nu} &=& \frac{\btil^\mu \btil^\nu}{\btil^2}, \quad
	\Pcal_4^{\mu\nu} = \frac{1}{\sqrt{\util^2 \btil^2}} \left( \util^\mu \btil^\nu + \util^\nu \btil^\mu \right), \label{eq.Q}
\end{eqnarray}
\begin{eqnarray}
	\Pi_1 &=& \frac{1}{\util^2} u_\mu u_\nu \Pi^{\mu\nu}, \quad
	\Pi_2 = g_{\mu\nu} \Pi^{\mu\nu} - \Pi_1 - \Pi_3, \nonumber \\
	\Pi_3 &=& \frac{1}{\btil^2} \left[ b_\mu b_\nu \Pi^{\mu\nu} + \frac{(b\cdot\util)^2}{\util^2} \Pi_1 - 2 \frac{b\cdot\util}{\util^2} u_\mu b_\nu \Pi^{\mu\nu} \right], \label{FF1} \\
	\Pi_4 &=& \frac{1}{\sqrt{\util^2 \btil^2}} \left( u_\mu b_\nu \Pi^{\mu\nu} - (b\cdot\util) \Pi_1 \right). \label{FF2}
\end{eqnarray}
In Eqs.~\eqref{eq.Q}--\eqref{FF2}, the vectors are defined as
\[
\util^\mu = u^\mu - \frac{q \cdot u}{q^2} q^\mu, \quad 
\btil^\mu = b^\mu - \frac{q \cdot b}{q^2} q^\mu - \frac{b \cdot \util}{\util^2} \util^\mu,
\] 
where 
$
b^\mu = \frac{1}{2B} \varepsilon^{\mu\nu\alpha\beta} F^\text{ex}_{\nu\alpha} u_\beta,
$
with $F^\text{ex}_{\nu\alpha}$ being the electromagnetic field strength tensor representing the external magnetic field.  
In the local rest frame (LRF) of the medium, these reduce to
\[
u^\mu_\text{LRF} = (1, \bm{0}), \quad 
b^\mu_\text{LRF} = (0, \hat{\bm{z}}),
\] 
where $\hat{\bm{z}}$ is along the direction of the external magnetic field.  
Using the Lorentz decomposition, the full $\rho^0$ propagator in a magnetic background is then expressed as
\begin{eqnarray} 
	D^{\mu\nu}(T,eB) &=& \frac{\Pcal_2^{\mu\nu}}{q^2 - m_\rho^2 + \Pi_2} 
	+ \frac{(q^2 - m_\rho^2 + \Pi_1) \Pcal_3^{\mu\nu}}{(q^2 - m_\rho^2 + \Pi_3)(q^2 - m_\rho^2 + \Pi_1) - \Pi_4^2}  
	- \frac{\Pi_4 \Pcal_4^{\mu\nu}}{(q^2 - m_\rho^2 + \Pi_1)(q^2 - m_\rho^2 + \Pi_3) - \Pi_4^2} \nn \\
	&& + \frac{(q^2 - m_\rho^2 + \Pi_3) \Pcal_1^{\mu\nu}}{(q^2 - m_\rho^2 + \Pi_3)(q^2 - m_\rho^2 + \Pi_1) - \Pi_4^2}  
	- \frac{q^\mu q^\nu}{q^2 m_\rho^2}. \label{eq.Dbar.B}
\end{eqnarray}
Here, $m_\rho$ denotes the vacuum (bare) $\rho$-meson mass, while the temperature- and magnetic-field-dependent medium effects are incorporated through the thermomagnetic self-energy	$\Pi^{\mu\nu}(q;T,eB)$.
The imaginary part of Eq.~\eqref{eq.Dbar.B} is used to obtain the spectral function $\mathcal{A}(q;T,eB)$, which is then employed to calculate the dilepton production rate.
The analytic form of the $\rho^0$ self-energy is obtained using an effective field theory of hadrons, where the $\rho$-meson interacts with pions via the interaction Lagrangian density~\cite{Krehl:1999km}:
\begin{equation}
	\mathcal{L}_\text{int} = - g_{\rho\pi\pi} \, (\partial_\mu \bm{\rho}_\nu) \cdot \left( \partial^\mu \bm{\pi} \times \partial^\nu \bm{\pi} \right), \label{SELint}
\end{equation}
where $\bm{\rho}_\nu$ and $\bm{\pi}$ are the isovector fields representing the $\rho$-mesons and pions respectively.  
The coupling constant $g_{\rho\pi\pi} = 20.72~\text{GeV}^{-2}$ is fixed from the decay width $\Gamma_{\rho\to\pi\pi} = 155.8~\text{MeV}$.
In the present work, we consider the pion-loop contribution to the $\rho^0$ self-energy corresponding to the $\rho^0\rightarrow\pi^+\pi^-$ channel, which is the dominant decay mode of the $\rho$ meson in vacuum. This choice allows us to investigate the effects of Landau-level	quantization of charged pions on the $\rho^0$ spectral function and the	resulting dilepton emission. We note that interactions of the $\rho$ meson with nucleons and other hadronic resonances can provide additional in-medium contributions, which may modify the dilepton emission rate and the source-level anisotropy coefficients	$v_n$. Their quantitative inclusion is beyond the scope of the present work.
The corresponding Feynman diagram for the one-loop self-energy of $\rho^0$ is shown in Fig.~\ref{FDSE}. To calculate the thermomagnetic $\rho^0$ self-energy, we employ the real-time formalism (RTF) of finite-temperature field theory. In RTF, the self-energy, which is a two-point function, forms a $2 \times 2$ matrix, but only the $11$-component is needed for the calculation of the analytic self-energy~\cite{Mallik:2016anp}. It can be expressed as $\text{Im}~\Pi^{\munu}(q)=\text{tanh}\FB{\frac{|q^0|}{2T}}\text{Im}~\Pi^{\munu}_{11}(q)$ and $\text{Re}~\Pi^{\munu}(q)=\text{Re}~\Pi^{\munu}_{11}(q)$. Using the interaction Lagrangian in Eq.~\eqref{SELint}, the 11-component of the $\rho^0$ one-loop self-energy is obtained as~\cite{Ghosh:2019fet,Mondal:2023vzx}:
\begin{figure}[h]
	\includegraphics[scale=.3]{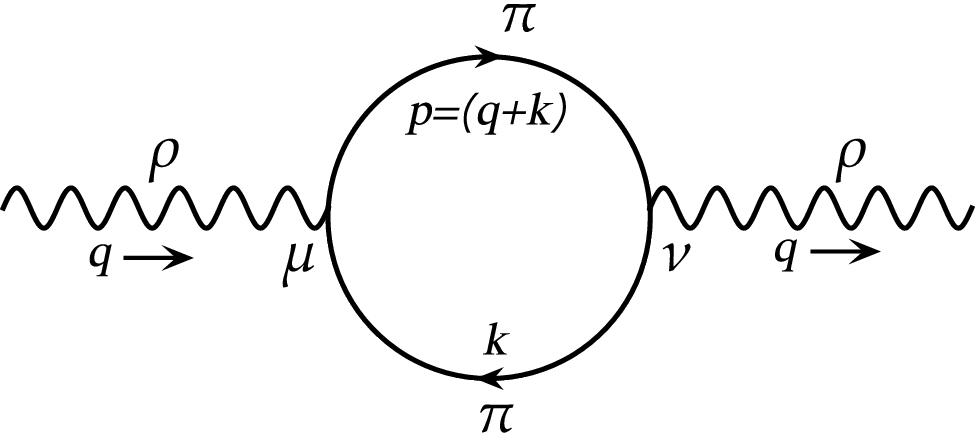}	
	\caption{Feynman diagram for the one-loop self-energy of the $\rho^0$ meson\cite{Mondal:2023vzx}.}\label{FDSE}
\end{figure} 
\begin{equation}
	\Pi^{\mu\nu}_{11}(q;T,eB) = i \int \frac{d^4 k}{(2\pi)^4} \, {\mathcal{K}^{\mu\nu}(q,k)} \, D^{\rm MagF}_{11}(k) \, D^{\rm MagF}_{11}(p=q+k), 
	\label{SEPi11}
\end{equation}
where {$\mathcal{K}^{\mu\nu}(q,k)$} contains the vertex factors and is given by:
\begin{equation}
	{\mathcal{K}^{\mu\nu}(q,k)} = g_{\rho\pi\pi}^2 \Big[ k^\mu k^\nu q^4 + q^\mu q^\nu (q \cdot k)^2 - (q^\mu k^\nu + q^\nu k^\mu)(q \cdot k) q^2 \Big]. \label{eq.N0}
\end{equation}
Here, $D^{\rm MagF}_{11}(p)$ is the $11$-component of the charged pion propagator, given by the expression~\cite{Ghosh:2019fet,Ayala:2016awt}:
\begin{equation}
	D^{\rm MagF}_{11}(p) = \sum_{l=0}^{\infty} 2(-1)^l L_l(2\alpha_p) e^{-\alpha_p} \Bigg[ 2 \pi i \, \eta(p \cdot u) \, \delta(p_\parallel^2 - m_l^2) + \frac{-1}{p_\parallel^2 - m_l^2 + i \epsilon} \Bigg], \label{DeltaPi2}
\end{equation}
where the sum runs over Landau levels $l$, $\alpha_p = -p_\perp^2 / eB$, and $m_l = \sqrt{m_\pi^2 + eB (2l+1)}$ is the effective pion mass.  
$L_l(z)$ denotes the Laguerre polynomial of order $l$, and $\eta(x) = \Theta(x) f_\text{BE}(x) + \Theta(-x) f_\text{BE}(-x)$, with $\Theta(x)$ being the Heaviside step function.  
The components of the momentum are defined as $p_{\parallel,\perp}^\mu = g_{\parallel,\perp}^{\mu\nu} p_\nu$ with 
$g_\parallel^{\mu\nu} = \text{diag}(1,0,0,-1)$ and $g_\perp^{\mu\nu} = \text{diag}(0,-1,-1,0)$.  
In this convention, $p_\parallel^2 = p_0^2 - p_z^2$ and $p_\perp^2 = -(p_x^2 + p_y^2)$.
Now, substituting Eq.~\eqref{DeltaPi2} into Eq.~\eqref{SEPi11} and performing the $dk^0$ integration, we obtain the real {part} $\rho^0$ self-energy as
\begin{eqnarray}
\text{Re}~\Pi^{\mu\nu}(q;T,eB)
	&=&
	\text{Re}~\Pi^{\mu\nu}_{\rm Vac}(q,eB)
	+ \sum_{n=0}^{\infty}\sum_{l=0}^{\infty}
	\int_{-\infty}^{\infty}\frac{dk_z}{2\pi}\,
	\mathcal{P}\Bigg[
	\frac{f_l^k}{2\omega_l^k}
	\Bigg\{
	\frac{N^{\mu\nu}_{nl}(q,k^0=-\omega_l^k,k_z)}
	{(q^0-\omega_l^k)^2-(\omega_n^p)^2}
	+
	\frac{N^{\mu\nu}_{nl}(q,k^0=\omega_l^k,k_z)}
	{(q^0+\omega_l^k)^2-(\omega_n^p)^2}
	\Bigg\}
	\nonumber \\
	&&\hspace{3.5cm}
	+
	\frac{f_n^p}{2\omega_n^p}
	\Bigg\{
	\frac{N^{\mu\nu}_{nl}(q,k^0=-q^0-\omega_n^p,k_z)}
	{(q^0+\omega_n^p)^2-(\omega_l^k)^2}
	+
	\frac{N^{\mu\nu}_{nl}(q,k^0=-q^0+\omega_n^p,k_z)}
	{(q^0-\omega_n^p)^2-(\omega_l^k)^2}
	\Bigg\}
	\Bigg].
	\label{eq.repi.B}
\end{eqnarray}
Here, $\mathcal{P}$ denotes the Cauchy principal value. In the above expressions, all $N^{\mu\nu}_{nl}$ are defined in Eq.~\eqref{Ncalmunu}. The quantity $\Pi^{\mu\nu}_{\rm Vac}(q,eB)$ represents the vacuum contribution in the presence of a magnetic field (see Refs.~\cite{Mondal:2023vzx,Ghosh:2019fet} for details):
\begin{equation}
	{\Pi^{\mu\nu}_{\rm Vac}(q, eB)=i~\sum_{n=0}^{\infty}\sum_{l=0}^{\infty}\int\frac{d^2k_{||}}{(2\pi)^2}~\frac{{N}^{\mu\nu}_{nl}(q,k_\parallel)}{k^2_{||}-m^2_l+i\epsilon}\,\frac{1}{(q_{||}+k_{||})^2-m^2_n+i\epsilon}},
\end{equation}
\begin{eqnarray}
{N}^{\mu\nu}_{nl}(q,k_\parallel)&=&\int\frac{d^2k_\perp}{(2\pi)^2}\,\tilde{\mathcal{K}}^{\mu\nu}_{nl}(q,k_\perp,k_\parallel),\label{Ncalmunu} \\
\tilde{\mathcal{K}}^{\mu\nu}_{nl}(q,k_\perp,k_\parallel)&=&4(-1)^{n+l} e^{-\alpha_k-\alpha_p}\,\mathcal{K}^{\mu\nu}\,L_l(2\alpha_k)\,L_n(2\alpha_p).
\end{eqnarray}
and $\omega_l^k=\sqrt{k_z^2 + m_l^2}$, $\omega_n^p=\sqrt{p_z^2 + m_n^2}$, $f_l^k=f_\text{BE}(\omega_l^k)$, $f_n^p=f_\text{BE}(\omega_n^p)$.  
\begin{figure}[h]	
	\includegraphics[angle = 0, scale=1.0]{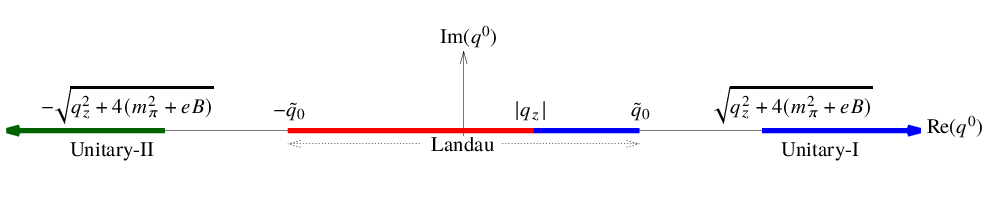}
	\caption{{(Color online) Analytic structure of the $\rho^0$ self-energy in the complex $q^0$ plane for the special case $\bm{q}=q_z\hat{z}$. Here, $\tilde{q}^0 = \sqrt{q_z^2 + \left(\sqrt{m_\pi^2 + eB} - \sqrt{m_\pi^2 + 3eB}\right)^2}$. The Unitary-I cut and a portion of the Landau cut (indicated by the blue line) correspond to the kinematically allowed region for physical dilepton emission. For a general momentum configuration $\bm{q}$, the gap between the Landau and Unitary-I cuts disappears, leading to a continuous dilepton emission spectrum ~\cite{Mondal:2023ypq}.}}
	\label{Fig.AS}
\end{figure}
\\
{Similarly, the imaginary part of the $\rho^0$ self-energy is given by}
\begin{eqnarray}
\text{Im}~\Pi^{\mu\nu}(q;T,eB)&=&-\tanh\!\left(\frac{|q^0|}{2T}\right)\pi \sum_{n=0}^{\infty}\sum_{l=0}^{\infty}\int_{-\infty}^{\infty}\frac{dk_z}{2\pi}\,\frac{1}{4\omega_l^k\,\omega_n^p}	\nonumber \\
	&&\hspace{-2cm}\times
	\Bigg[
		(1 + f_l^k + f_n^p + 2 f_l^k f_n^p)	\Big\{N^{\mu\nu}_{nl}(q,k^0=-\omega_l^k,k_z)\,\delta(q^0 - \omega_l^k - \omega_n^p)	+ N^{\mu\nu}_{nl}(q,k^0=\omega_l^k,k_z)\,\delta(q^0 + \omega_l^k + \omega_n^p)	\Big\}	\nonumber \\
		&&\hspace{-2cm}
		+(f_l^k + f_n^p + 2 f_l^k f_n^p)\Big\{	N^{\mu\nu}_{nl}(q,k^0=-\omega_l^k,k_z)\,\delta(q^0 - \omega_l^k + \omega_n^p) +	N^{\mu\nu}_{nl}(q,k^0=\omega_l^k,k_z)\,	\delta(q^0 + \omega_l^k - \omega_n^p) \Big\}
	\Bigg]
	\label{ImPi1}
\end{eqnarray}
The different delta-function structures in Eq.~(\ref{ImPi1}) correspond to distinct physical processes and define branch cuts of the self-energy in the complex energy plane (Fig.~\ref{Fig.AS} for the special case when $\bm{q}=q_z\hat{z}$):
	\begin{itemize}
		\item The terms proportional to $\delta(q^0 - \omega_l^k - \omega_n^p)$ and $\delta(q^0 + \omega_l^k + \omega_n^p)$ correspond to particle decay and production processes, and give rise to the {Unitary-I} and {Unitary-II} cuts, respectively. 
		\item The terms involving $\delta(q^0 - \omega_l^k + \omega_n^p)$ and $\delta(q^0 + \omega_l^k - \omega_n^p)$ correspond to scattering processes in the medium, leading to the {Landau cuts}. Unlike the $eB=0$ case, a nontrivial Landau-cut contribution can appear in the physical time-like region. This occurs when the pions in the loop occupy different Landau levels.
	\end{itemize}
Further, the imaginary part of the self-energy can be simplified by performing the $dk_z$ integration, yielding
\begin{eqnarray}
	\text{Im}~\Pi^{\mu\nu}(q;T,eB) &=& -\tanh\left(\frac{|q^0|}{2T}\right) 
	\sum_{n=0}^{\infty} \sum_{l=0}^{\infty} \frac{1}{4 \lambda^{1/2}(q_\parallel^2, m_l^2, m_n^2)}
	\sum_{k_z \in \{k_z^\pm\}} 
	\Bigg[ (1 + f_l^k + f_n^p + 2 f_l^k f_n^p) \nonumber \\
	&& \times \Big\{ 
	\Theta \Big(q^0 - E^{nl}_U\Big) \, N^{\mu\nu}_{nl}(q,k^0=-\omega_l^k,k_z) 
	+ \Theta \Big(-q^0 - E^{nl}_U \Big) \, N^{\mu\nu}_{nl}(q,k^0=\omega_l^k,k_z) 
	\Big\} \nonumber \\
	&& + (f_l^k + f_n^p + 2 f_l^k f_n^p) \Big\{
	\Theta \Big(q^0 - E^{\text{min}}_{L} \Big) \, 
	\Theta \Big(-q^0 + E^{\text{max}}_{L}\Big) \, N^{\mu\nu}_{nl}(q,k^0=-\omega_l^k,k_z) \nonumber \\
	&& + \Theta \Big(-q^0 - E^{\text{min}}_{L}\Big) \, 
	\Theta \Big(q^0 + E^{\text{max}}_{L}\Big) \, N^{\mu\nu}_{nl}(q,k^0=\omega_l^k,k_z) 
	\Big\} 
	\Bigg].
	\label{ImPi3}
\end{eqnarray}
Here, $k_z^\pm=\frac{-\tilde{q}_\parallel^2q_z\pm|q^0|\lambda^{1/2}(q_\parallel^2,m_l^2,m_n^2)}{2 q_\parallel^2}$,  
$\tilde{q}_\parallel^2 = q_\parallel^2 + m_l^2 - m_n^2$, K\"all\'en function $\lambda(x,y,z) = x^2 + y^2 + z^2 - 2xy - 2yz - 2zx$,~$E^{nl}_{U}=\sqrt{q_z^2 + (m_l + m_n)^2}$,~$E^{\text{max(min)}}_{L}=\text{max(min)}(q_z, E^\pm)$ with $E^\pm = \frac{m_l - m_n}{|m_l \pm m_n|} \sqrt{q_z^2 + (m_l \pm m_n)^2}$. In particular $E^{nl}_U$ and $E^{\text{max(min)}}_{L}$ in Eq.~\eqref{ImPi3} respectively denote Unitary and Landau cut thresholds for a particular $(n,l)$.
The Unitary-I and Unitary-II contributions are non-vanishing in the kinematic regions
$
\sqrt{q_z^2 + 4(m_\pi^2 + eB)} < q^0 < \infty$ and $-\infty < q^0 < -\sqrt{q_z^2 + 4(m_\pi^2 + eB)}
$
respectively, while the Landau-cut contributions are nonzero in the region
$
|q^0| < E^{\text{max}}_{L}$.
The Unitary-I and Landau cuts encode kinematically allowed decay and scattering processes of the $\rho^0$ meson in a magnetized medium~\cite{Ghosh:2019fet,Mondal:2023vzx}. 
The analytic structure becomes clearer when $\bm{q} = q_z \hat{z}$. In this case, Eq.~\eqref{ImPi3} simplifies, as $l = n-1,\, n,\, n+1$ for a given $n$, due to the structure of $N^{\mu\nu}_{nl}$, which can be expressed as a linear combination of $\delta_{l}^{n-1}$, $\delta_{l}^{n}$, and $\delta_{l}^{n+1}$ with appropriate coefficients (see Ref.~\cite{Mondal:2023vzx}). This choice of $\bm{q}=q_z\hat{z}$ leads to a clear separation between the Unitary and Landau cuts, as shown in Fig.~\ref{Fig.AS}. 
However, for a general momentum configuration $\bm{q}$ considered in our numerical results, the dilepton production rate is continuous spectrum and such a clear separation between the cuts does not occur.

\section{Numerical Results}\label{Numerical}
In this section, we present numerical results for the {source-level} dilepton production rate and its azimuthal angle dependence as a function of invariant mass $ M $, transverse momentum $ q_T $ with respect to beam axis (see Fig.~\ref{Frame}) {corresponding to vanishing longitudinal momentum in the chosen frame} $( q_x = 0 )$ from a magnetized hadronic matter. {At present, a direct comparison with experimentally measured observables is not available.} 
The numerical evaluation of the dilepton production rate involves a summation over Landau levels. In practice, this infinite sum is truncated at a finite upper limit. {For the results presented here, we include up to 500 Landau levels, which ensures numerical convergence for the range of magnetic field strengths and kinematic variables considered in this work. We have explicitly verified that increasing the cutoff does not lead to any significant change in the results within the relevant kinematic domain. We note, however, that the number of Landau levels required for convergence can, in general, depend on the value of $eB$ as well as the external kinematics.}
Since we are interested for hadronic stage of the matter created in HICs, we have chosen two representative values of temperature, $ T = 130 $ and $ 160 $ MeV respectively. As the hadrons are created in the later stages of the evaluation processes in HICs, it is expected that the strength of background field should be smaller compared to the QGP phase. Hence, we have chosen the lowest value of magnetic field $ eB = 0.01 $ GeV$ ^2 $ and a {higher} value of magnetic field $ eB = 0.05 $ GeV$ ^2 $ to understand the interplay between magnetic field and thermal effects.  
Since the magnetic field is oriented along the $z$-axis, it breaks rotational symmetry in the plane perpendicular to the beam direction. Consequently, the dilepton production rate (DPR) is expected to exhibit azimuthal-angle dependence, i.e., anisotropy in the transverse plane relative to the beam. To investigate this, we analyze various {anisotropy}  coefficients of the dilepton emission under different physical conditions.

The numerical evaluation of the integral {in the denominator} of Eq.~\eqref{Vn} can be simplified employing spatial symmetries of the magnetized matter {in the present setup}~\cite{Wang:2022jxx}. Firstly, we note that, with our choice of axes, the system remains invariant under spatial rotation {about the $ z $ axis}, which is a subgroup of the spatial rotation group that remains unbroken in $ B\ne 0 $ case. Invoking this symmetry we can write \[\frac{dN}{d^4xd^4q}(\pi-\phi)=\frac{dN}{d^4xd^4q}(\phi).\] Again, since the magnetic field is an axial vector, there is also a reflection symmetry with respect to the reaction plane. This will allow us to write \[\frac{dN}{d^4xd^4q}(\phi) = \frac{dN}{d^4xd^4q}(-\phi).\] Hence, we need to evaluate the integral only in the first quadrant, i.e. $ 0 \le \phi  < \pi/2 $. The results for the other three quadrants can be evaluated using the above symmetries.
\begin{figure}[h]	
	\includegraphics[angle = -90, scale=0.34]{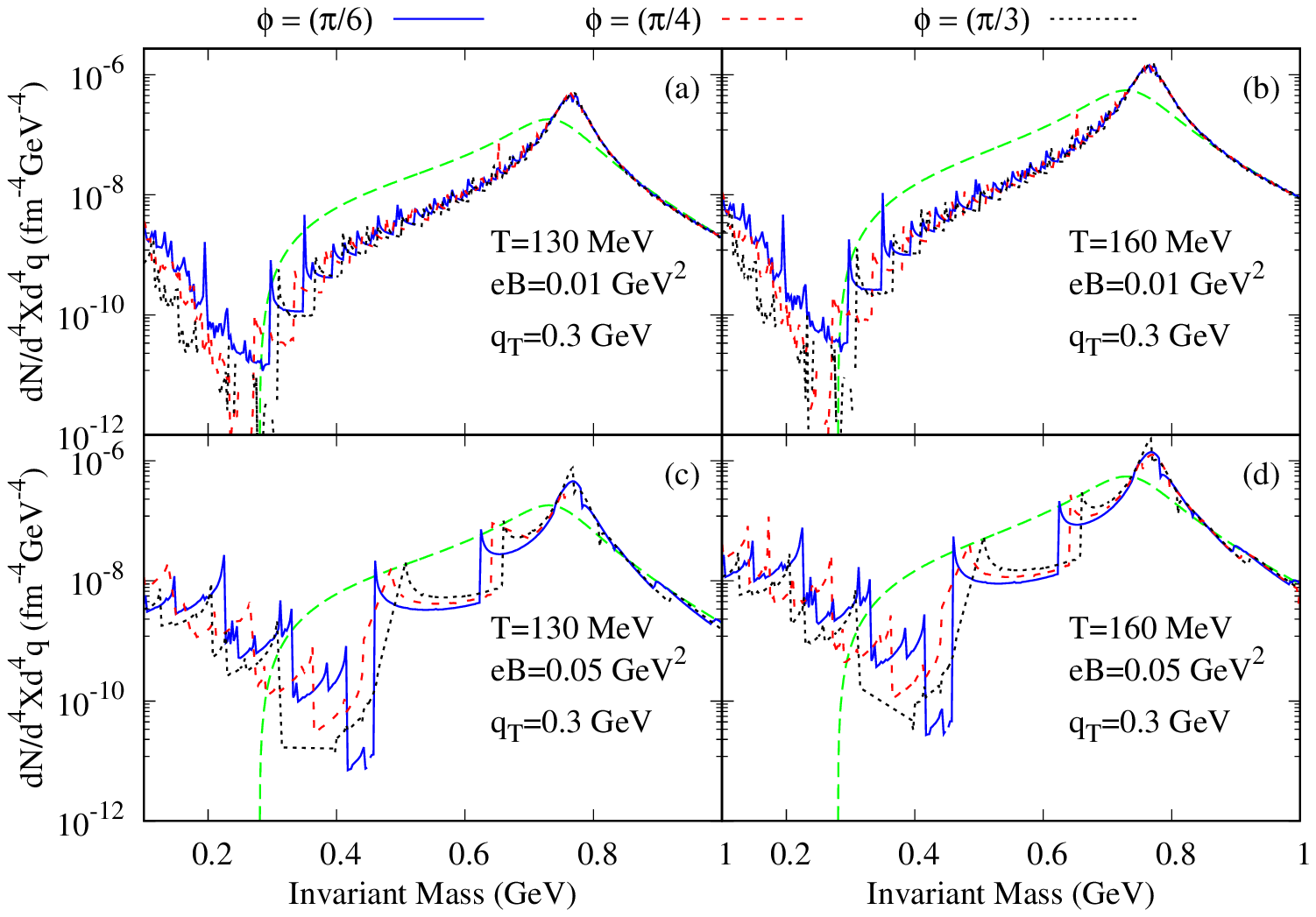}~~~~~~~
	\includegraphics[angle = -90, scale=0.34]{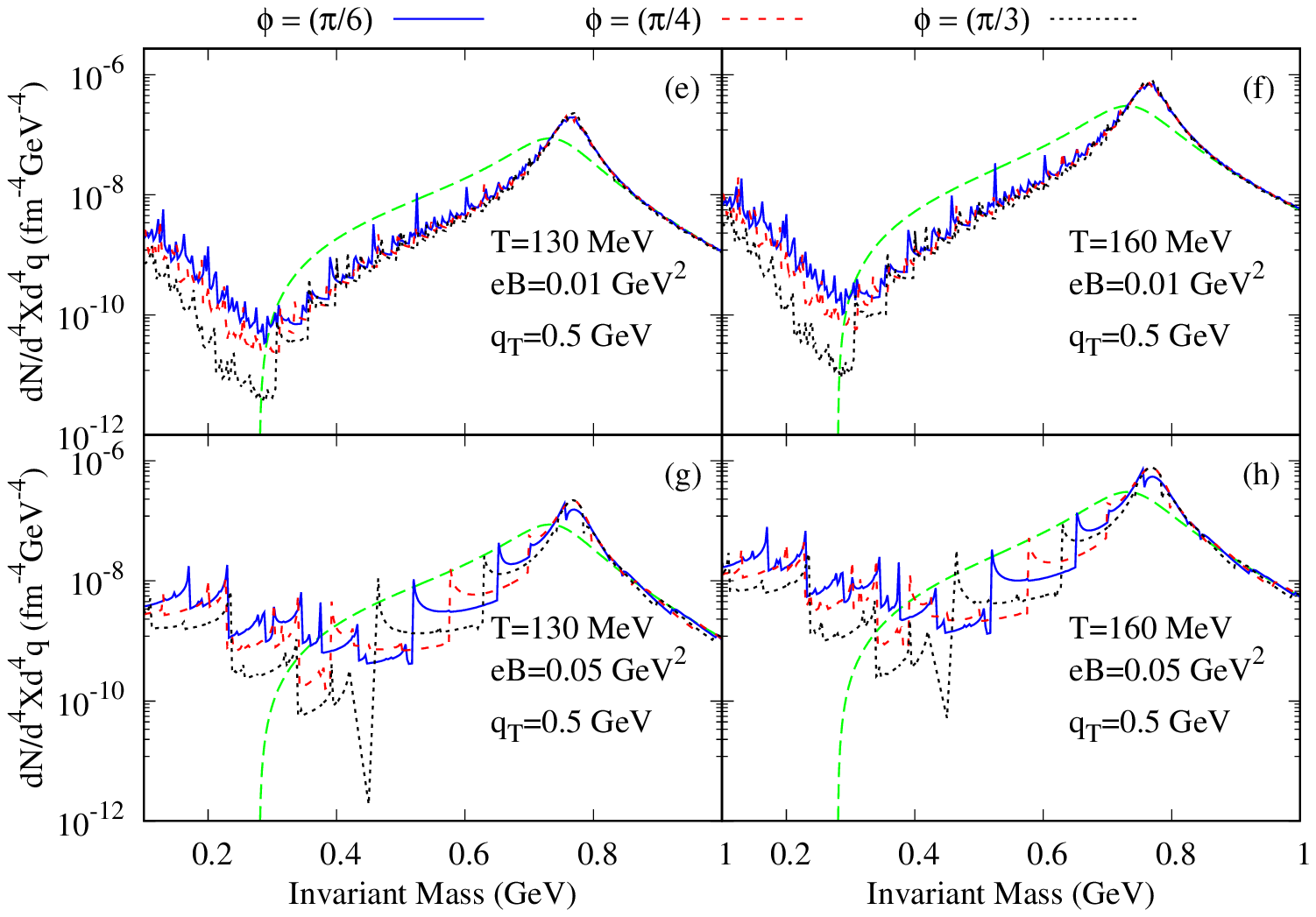}
\caption{(Color online) The dilepton production rate as a function of invariant mass for $eB=0.01,~0.05 \rm~GeV^2$ and $T=130,~160\rm~MeV$ at azimuthal angle $\phi=\frac{\pi}{6},\frac{\pi}{4},~\frac{\pi}{3}$ in (a)-(d) for $q_T=0.3\rm GeV$ and (e)-(h) for $q_T=0.5\rm GeV$. Green color line corresponds Born Rate $(eB=0)$.}
	\label{Fig.DPR}
\end{figure}

Figs.~\ref{Fig.DPR}(a)-(d) for $q_T=0.3$ GeV and Figs.~\ref{Fig.DPR}(e)-(h) for $q_T=0.5$ GeV show the variation of the dilepton production rate as a function invariant mass for different values of azimuthal angle $\phi$, magnetic field $eB$ and temperature $T$. For comparison, the born rate in absence of the background field (green dashed line) have also shown which is consistent with the earlier observations by C. Gale and J. Kapusta in Refs.~\cite{Gale:1988vv,Gale:1990pn}. Figs.~\ref{Fig.DPR}(a)-(d) as well as Figs.~\ref{Fig.DPR}(e)-(h) contain a set of four panels in which $T$ increases in the horizontal direction and $eB$ increases in the downward direction. To begin the discussion, we focus on Figs.~\ref{Fig.DPR}(a)-(d). It is evident from the plots that the dilepton emission rate is highly enhanced in the low invariant mass region compared to the born rate(in absence of $eB$) owing to the Landau cut contributions. This is a purely magnetic field dependent effect and can be connected to the process where $\rho$-meson is absorbed by means of scattering with a pion in lower Landau level and producing a pion in higher Landau level in the final state (and the time reversed process) because pion mass is modified by the external magnetic field, given by the expression $m_l=\sqrt{m^2_\pi+eB(2l+1)}$ (see below of Eq.~\eqref{DeltaPi2}). The last two terms of Eq.~\eqref{ImPi3} include Landau cut contributions. The DPR in higher invariant mass region is dominated by the unitary cut contributions corresponding to the decay process $\rho_0\to\pi^+\pi^-$. These can be connected to the first two terms of Eq.~\eqref{ImPi3}. A detailed study regarding of these different cuts structure of kinematic domain can be found in Ref.~\cite{Ghosh:2019fet,Mondal:2023vzx}. Moreover, because the lepton masses are negligible compared to the pion mass, the threshold invariant mass for dilepton production aligns with the $\rho$-meson mass for all magnetic-field strengths and temperatures. Furthermore, an increase in temperature leads to a noticeable rise in the overall magnitude of the DPR as evident from the comparison between Figs.~\ref{Fig.DPR}(a) and ~\ref{Fig.DPR}(b). This behavior stems from the enhanced thermal phase space at higher temperatures. It is {worth mentioning} that when $q_T$ is along the $z$-direction, the Landau and Unitary cut contributions are separated, creating a forbidden gap that is independent of temperature and the background magnetic field~\cite{Mondal:2023vzx}. This arises because a pion in Landau level $n$ can interact with pions in levels $n-1$, $n$, or $n+1$ to produce a $\rho$-meson. In contrast, in our study, $q_T$ points in a different or arbitrary direction, resulting in a continuous dilepton emission spectrum.
To study the azimuthal angle dependence of DPR, we {have chosen} representative values of $\phi=$$\frac{\pi}{3},~\frac{\pi}{4},~\frac{\pi}{6}$ as the knowledge of dilepton emission in first quadrant is sufficient to evaluate the results in the other three quadrants. It is clear from the figures that dilepton production rate has a non-trivial azimuthal angle dependence in the low invariant mass region. As the magnetic strength increases, the $\phi$-dependence qualitatively increases as comparing between Figs.~\ref{Fig.DPR}(a) and (c). On the other hand, the rate has negligible $\phi$ dependence for higher values of invariant mass of dilepton  indicating an isotropic emission of high mass dileptons.
The spikes in the DPR arise from “threshold singularities” which appear whenever the energy of the virtual photon becomes equal to one of the numerous Landau-level thresholds. This can be understood from Eq.~\eqref{ImPi3}, where the Källén function in the denominator vanishes at these thresholds, as defined by the unit step function. { While care has been taken to control numerical uncertainties, these features (spikes and discontinuities) are physical in origin and reflect the discrete Landau-level structure of the charged pion propagators in a magnetic field}.
However, this non-smooth nature of DPR spectrum do not lead to any generic trend in azimuthal angle dependence. Finally, for a given value of $eB$ and $T$, the overall DPR is suppressed with larger value of $q_T$ which is evident by comparing Fig.~\ref{Fig.DPR}(a) and Fig.~\ref{Fig.DPR}(e). It is connected with the modification of effective temperature of the medium. Therefore, observed thermal suppression is primarily due to the Bose-Einstein distribution appearing in the DPR expression in Eq.~\eqref{DPR}. Further reduction can be attributed to the spectral function or equivalently the imaginary part of the $\rho^0$ self-energy (see Appendix), which also involves Bose-Einstein distribution function. However, obtaining a quantitative estimate requires a complete space-time evolution of this rate within magneto-hydrodynamics, which lies beyond the scope of the present study. In addition, the magnitude of {emission} rate increases with the increasing temperature due to increase of the thermal phase space.

\begin{figure}[h]	
	\includegraphics[angle = -90, scale=0.35]{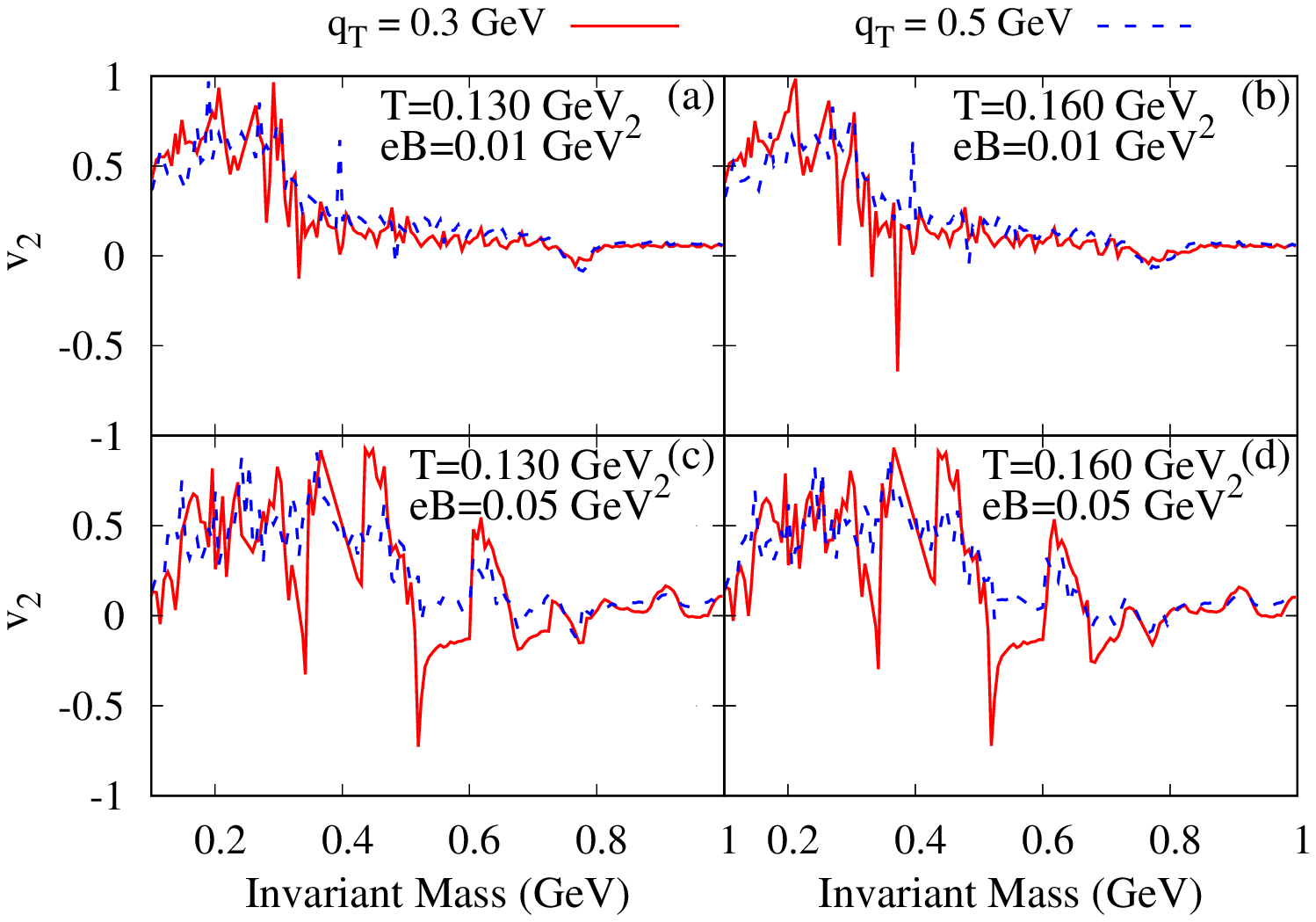}~
	\includegraphics[angle = -90, scale=0.35]{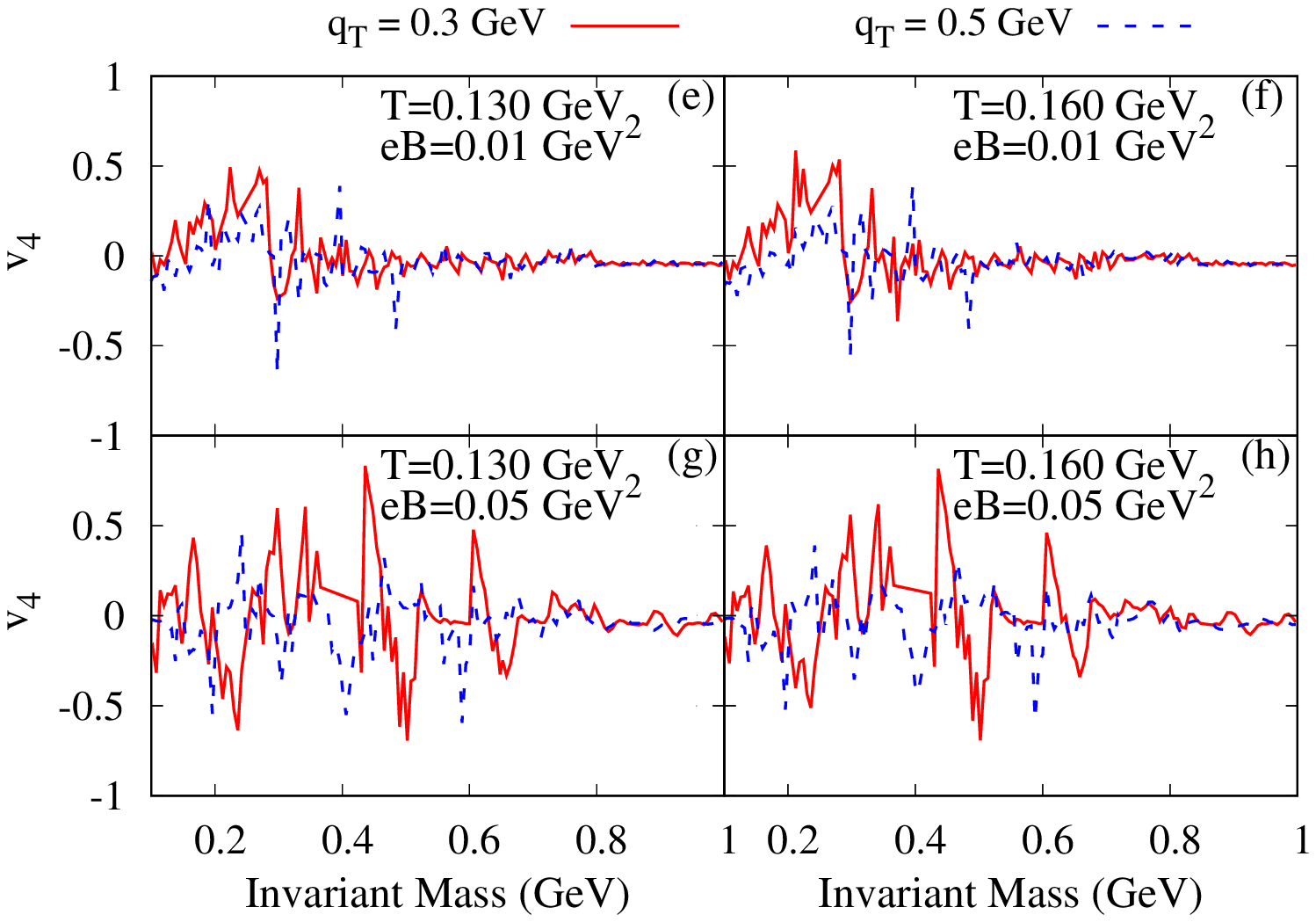}
	\includegraphics[angle = -90, scale=0.35]{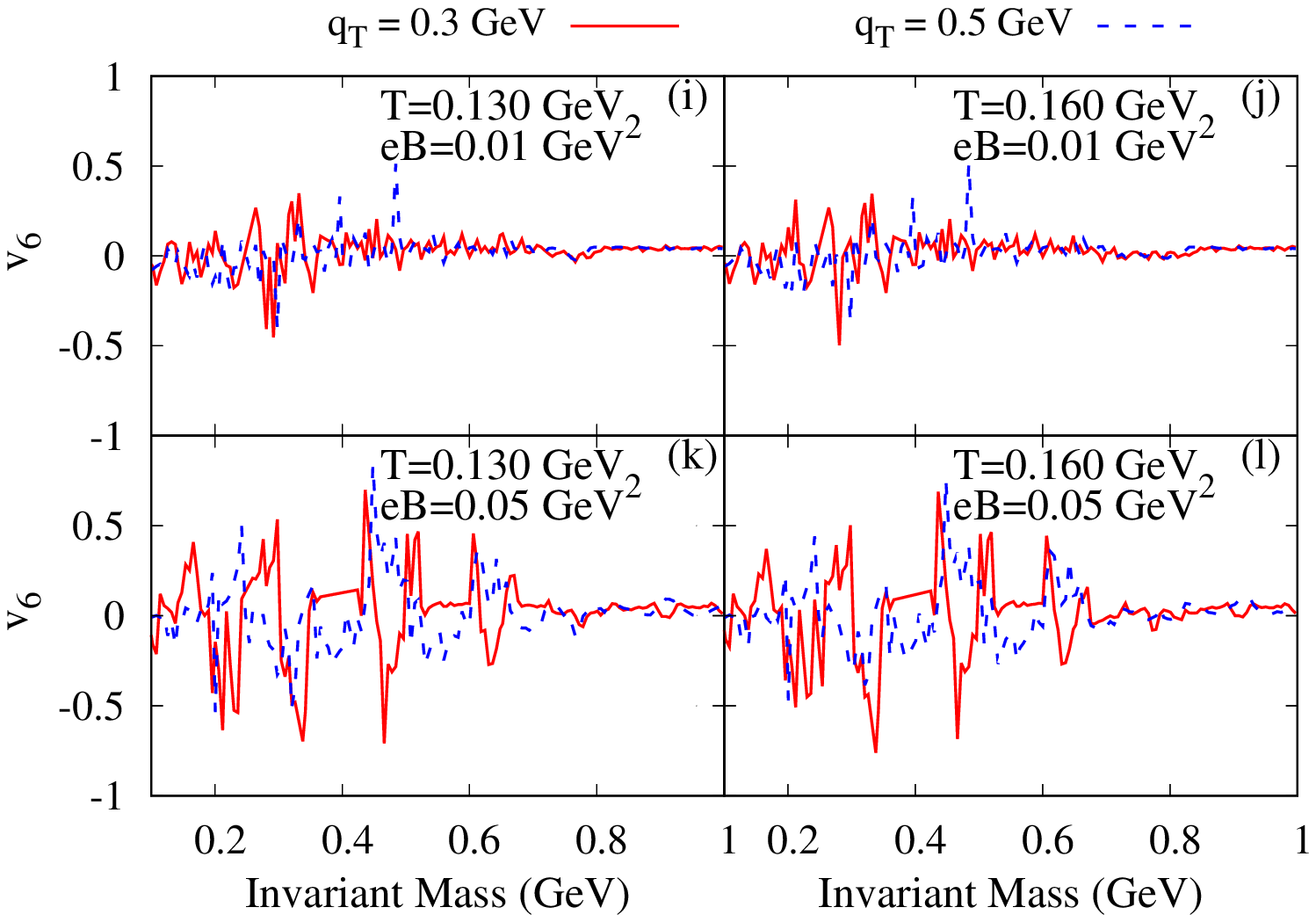}
	\caption{(Color online) The source-level azimuthal anisotropy coefficients $v_n$ as a function of invariant mass $M$ for $v_2$ in panels (a)--(d), $v_4$ in panels (e)--(h), and $v_6$ in panels (i)--(l), for $eB=0.01,\,0.05~\mathrm{GeV}^2$ and $q_T=0.3,\,0.5~\mathrm{GeV}$ at	$T=130$ and $160~\mathrm{MeV}$. The results are obtained by separately binning the numerator and denominator of $v_n$ over the invariant-mass intervals.}
	\label{Fig.Vn}
\end{figure}
As discussed earlier, the anisotropy in dilepton production can be quantified by analyzing flow coefficients $(v_n)$ defined in Eq.~\eqref{Vn}. The odd flow coefficients (particularly $v_1$, $v_3$) vanish because the dilepton emission rate is symmetric under the transformation $\phi\to\pi-\phi$ as discussed in the {beginning} of this section. Hence, we only analyze the even flow coefficients for dilepton emission in this section. 
To reduce numerical fluctuations associated with the sharp	Landau-level threshold structures, we bin the numerator and denominator	of $v_n$ separately over the invariant-mass intervals and subsequently take their ratio to obtain the binned anisotropy coefficients. 
Figs.~\ref{Fig.Vn}(a)-(d) show variation of the ellipticity parameter $(v_2)$ as a function of invariant mass for different transverse momenta in the similar setting namely in the horizontal (downward) direction the temperature (background magnetic field) increases keeping $eB(T)$ constant. The numerical data reveal various qualitative features, which are also influenced by the emergence of several spike-like structures associated with numerous Landau level quantization and numerous threshold effects. The figure shows that $v_2$ has a strong tendency to remain positive in low invariant mass region. 
A positive value of $v_2$ reflects an oblate shape of the DPR, indicating enhanced dilepton production around $\phi\sim0$, i.e., in a	direction transverse to the background magnetic field, as shown in Fig.~\ref{Frame}.
Furthermore, $v_2$ displays {oscillations} increasing with the increase of magnetic field at low invariant mass region and the oscillations become negligible at high invariant mass region suggesting that the DPR becomes nearly isotropic, a trend also evident in the earlier figures Figs.~\ref{Fig.DPR}(a)-(h). The figure ( for example Fig.~\ref{Fig.Vn}(a) ) also indicates a general upward trend of $v_2$ with increasing $q_T$. The $q_T$ dependence on $v_2$ is more evident with the large value of magnetic field as observed from Figs.~\ref{Fig.Vn}(a) and (c) or Figs.~\ref{Fig.Vn}(b) and (d). 
The results for $v_2$ in a magnetized hadronic medium have been reported previously in Ref.~\cite{Mondal:2023ypq}. The present results also show qualitative similarities with the calculation for a magnetized quark-gluon plasma reported in Ref.~\cite{Wang:2022jxx}. 
Now, we extend our analysis to higher-order anisotropy coefficients $v_4$ and $v_6$. {It probes finer} details of the medium in heavy-ion collisions. We will concentrate our attention on the same kinematic region. The numerical results for dilepton $v_4$ are shown in Figs.~\ref{Fig.Vn}(e)-(h). As before, the four panels show results under identical conditions: temperature (or magnetic field) increases horizontally (or downward) with constant $eB(T)$. It is observed that the coefficient $v_4$ tends to be positive at low invariant masses, indicating significant anisotropy in the DPR. However, the value of $v_4$ becomes negligible at higher invariant masses. Here, also $q_T$ dependence of $v_4$ is prominent with increasing magnetic field. The results for dilepton $v_6$ are presented in Figs.~\ref{Fig.Vn}(i)-(l) with four panels, temperature (or magnetic field) increases horizontally (or downward) with constant $eB(T)$. The dilepton $v_6$ fluctuates at low invariant mass but remains nearly zero at higher invariant masses. Similar to $v_2$ and $v_4$, the $q_T$ dependence on $v_6$ significantly increased with increasing magnetic field. Notably, we observe well-pronounced modulations in the $v_n$ dependence on the invariant mass. In addition to the alternating sign pattern, the high-resolution data exhibit other intriguing features. A comparison of the results in Figs.~\ref{Fig.Vn}(a)--(l) reveals correlated peak patterns across all anisotropy coefficients, particularly in the low invariant mass region. In addition, the coefficients does not show any strong dependence on temperature.
The oscillations in the anisotropy coefficients $v_n$ arise from the so-called threshold singularities associated with the	Landau-level quantization of pions. In the present calculation, no additional finite thermal or collisional width is introduced for the	pion or $\rho$ propagators. In a realistic hadronic medium, finite thermal widths and collisional broadening would smear these sharp threshold structures. Therefore, the spikes observed in $v_n$ should not be interpreted as experimentally observable features, but rather as signatures of the underlying Landau-level
threshold contributions.

\section{Summary \& Conclusion}\label{SC}
Our study demonstrates that a background magnetic field significantly modifies the {local or source-level} dilepton production rate and its azimuthal anisotropy in a hot hadronic medium. The key findings--enhanced low-mass dilepton production due to Landau-cut contributions, strong azimuthal anisotropy quantified by oscillatory behavior of the source-level coefficients $(v_n)$ in the same mass region and isotropic emission at higher invariant masses--provide a clear picture of magnetically induced effects in a static medium.

{These results can be contextualized within the broader literature on electromagnetic probes in heavy-ion collisions. The enhancement of the DPR is consistent with prior theoretical works in the presence of magnetic fields~\cite{Ghosh:2019fet,Hattori:2012je}. The key quantity governing the DPR is the thermomagnetic in-medium spectral function of the $\rho^0$ meson, i.e., the imaginary part of the full interacting $\rho^0$ propagator, evaluated using the real-time formalism of thermal field theory and the Schwinger proper-time method. 	
The analytic structure in the complex energy plane indicates that, in addition to the conventional contribution from the Unitary cut beyond the two-pion threshold, a non-trivial Landau-cut contribution also appears in the physical kinematic domain. This arises because the charged pions involved in the loop can occupy different Landau levels before and after scattering with the $\rho^0$ meson. 	
In our calculation, the Landau-cut contribution is a purely magnetic phenomenon, i.e, absent at zero magnetic-field and dominates in the low invariant-mass region. In contrast, the Unitary cut governs the high invariant-mass region where the sensitivity to the magnetic field becomes negligible. 
}
The most significant implication of our work lies in the quantification of intrinsic anisotropy. The positive and oscillatory $v_2$, along with non-zero higher-order coefficients ($v_4$, $v_6$), at low invariant masses establish that a static, magnetized medium itself is an anisotropic source of dilepton emission.  
This finding is crucial because it suggests that the azimuthal anisotropy coefficients ($v_n$) measured in heavy-ion collisions likely receive contributions from two distinct sources: (i) the collective hydrodynamic flow of the medium, and (ii) the intrinsic anisotropy induced by the strong, early-stage magnetic field. 
{However, it is important to emphasize that the anisotropy coefficients defined in this work correspond to the local (source-level) emission rate and should not be directly identified with experimentally measured flow coefficients. In relativistic heavy-ion collisions, the observed anisotropic flow arises predominantly from the collective hydrodynamic response of the medium to initial spatial anisotropies, and is obtained after integrating the local emission rate over the full space-time evolution of the system. Consequently, the anisotropies reported here represent intrinsic emission properties, which may be significantly modified—or even washed out—by the subsequent dynamical evolution of the medium.}

However, several limitations of this study must be acknowledged. First, our model assumes a constant, uniform magnetic field in a static, infinite medium. In an actual heavy-ion collision, the magnetic field is extremely transient, decaying rapidly with time, and is highly inhomogeneous. Second, we have not considered the concurrent evolution of the medium --- its expansion, cooling, and the gradual dilution of the magnetic field. These dynamic aspects are critical for direct phenomenological comparison with experimental data. Third, the hadronic medium is treated in a simplified manner; potential effects from other hadronic states or a phase transition to a quark-gluon plasma are not included.

Given these limitations, the logical direction for future research is clear. To bridge the gap between our static, magnetized calculation and the dynamic reality of heavy-ion collisions, detailed numerical simulations are essential. Future work should employ phenomenological frameworks, such as relativistic hydrodynamic or transport simulations, that self-consistently incorporate both the spacetime evolution of the hot medium and that of the embedded electromagnetic fields. Such simulations could convolve the intrinsic anisotropic emission rates calculated here with a realistic fireball evolution to produce predictions for the final observed $v_n$ coefficients. This would allow for a quantitative separation of hydrodynamic and magnetic contributions to dilepton anisotropy.

In conclusion, our findings provide strong qualitative evidence that a background magnetic field acts as an intrinsic source of azimuthal anisotropy in dilepton emission. While definitive claims regarding the use of this anisotropy as a clean signature of primordial magnetic fields in experiments require the aforementioned sophisticated, integrated simulations, this work establishes a foundational mechanism and provides benchmark results for such future endeavors.
{We also note that the present analysis is performed in the rest frame of a static medium with a fixed magnetic-field direction. In realistic heavy-ion collisions, the medium undergoes collective expansion, and the local rest frame of each fluid element generally differs from the laboratory frame. A realistic comparison with experimental data would therefore require a covariant extension of the present results incorporating flow effects and Lorentz transformations of both the dilepton momentum and the electromagnetic field. Therefore, the results presented here should be interpreted as providing qualitative insight into magnetic-field effects on dilepton production and possibly as an upper bound on related observables.}

\section{ACKNOWLEDGMENT}
Authors thank Prof. Sourav Sarkar, Prof. Pradip Roy, Dr. Nilanjan Chaudhuri, and Dr. Snigdha Ghosh for valuable discussions at various stages of this work. 

\bibliographystyle{apsrev4-1}
\bibliography{Ref}

\end{document}